\documentclass{kapproc} 

\RequirePackage{graphicx}
\RequirePackage{epsf}
\input{psfig.sty}

\upperandlowercase
\let\footnote\savefootnote
\let\footnotetext\savefootnotetext 
\let\footnoterule\savefootnoterule 
\kluwerbib 

\begin{document}


\articletitle{The Initial Conditions to Star Formation: Low Mass Stars
at Low Metallicity}


\chaptitlerunninghead{Low Mass Stars at Low Metallicity}



 \author{Martino Romaniello}
 \affil{European Southern Observatory, Karl-Schwarzschild-Strasse 2,
  D-85748 Garching bei M\'unchen, Germany}
 \email{mromanie@eso.org}

  \author{Nino Panagia, and Massimo Robberto} \affil{Space telescope
  Science Institute, 3700 San Martin Drive, Baltimore, MD 21218}
  \email{panagia@stsci.edu,robberto@stsci.edu}





\begin{abstract}
We have measured the present accretion rate of roughly 800 low-mass
($\sim1-1.4~M_\odot$) pre-Main Sequence stars in the field of SN~1987A
in the Large Magellanic Cloud. The stars with statistically
significant Balmer continuum and Halpha excesses are measured to have
accretion rates larger than $\sim 1.5\times10^{-8}M_\odot\ yr^{-1}$ at
an age of 12-16~Myrs. For comparison, the time scale for disk
dissipation observed in the Galaxy is of the order of 6~Myrs.
\end{abstract}

\section{Introduction}
From an observational standpoint, most of the effort to characterize
and understand the process of star formation has traditionally been
devoted to nearby Galactic star-forming regions, such as the
Taurus-Auriga complex, Orion, etc. If this, on the one hand, permits
one to observe very faint stars at the best possible angular
resolution, on the other it is achieved at the expense of probing only
a very limited set of initial conditions for star formation (all these
clouds have essentially solar metallicity, e.g., Padget 1996).
Studying the effects of a lower metallicity on star formation is also
essential to understand the evolution of both our own Galaxy, in which
a large fraction of stars were formed at metallicities below solar,
and what is observed at high redshifts. With a mean metallicity of
about a third solar, the Large Magellanic Cloud (LMC) provides an
ideal environment for these kinds of studies:

\begin{itemize}
\item with a distance modulus of $18.57\pm0.05$ (see the discussion in
  Romaniello et al 2000), the LMC is our closest galactic companion
  after the Sagittarius dwarf galaxy. At this distance one arcminute
  corresponds to about 15~pc and, thus, one pointing with a typical
  imaging instrument comfortably covers almost any star forming region
  in the LMC (10~pc see, e.g., Hodge 1988);
\item the depth of the LMC along the line of sight is
  negligible, at least in the central parts we consider (van der Marel
  \& Cioni 2001). All of the stars can, then, effectively be considered
  at the same distance, thus eliminating a possible spurious scatter in
  the Color-Magnitude Diagrams;
\item he extinction in its direction due to dust in our Galaxy is
  low, about $\mathrm{E(B-V)}\simeq 0.05$ (Bessell 1991) and, hence,
  our view is not severely obstructed.
\end{itemize}

There is currently a widespread agreement that low mass stars form by
accretion of material until their final masses are reached
(e.g., Bonnell et al 2001 and references therein). As a consequence,
the accretion rate is arguably \emph{the} single most important
parameter governing the process of low-mass star formation and its
final results, including the stellar Initial Mass Function.  Ground
and HST-based studies show that there may be significant differences
between star formation processes in the LMC and in the Galaxy. For
example, Lamers et al (1999) and de Wit et al (2002) have identified
by means of ground-based observations high-mass pre-Main Sequence
stars (Herbig AeBe stars) with luminosities systematically higher than
observed in our Galaxy, and located well above the ``birthline'' of
Palla \& Staler (1990). They attribute this finding either to a
shorter accretion timescale in the LMC or to its smaller dust-to-gas
ratio. Whether such differences in the physical conditions under which
stars form will generally lead to differences at the low mass end is
an open question, but Panagia et al (2000) offer tantalizing evidence
of a higher accretion also for LMC low mass stars.

In this contribution we present the first measurement of the accretion
rate onto low-mass pre-Main Sequence stars outside of our Galaxy. The
full details of our analysis are reported in Romaniello et al (2004).

\section{Measuring the accretion rate}
The field of SN~1987A in the LMC was repeatedly imaged over the years
with the WFPC2 on-board the HST to monitor the evolution of its
Supernova remnant. We have taken advantage of this wealth of data and
selected from the HST archive a uniform dataset providing broad-band
coverage from the ultraviolet to the near infrared, as well as imaging
in the $\mathrm{H}\alpha$ line.

The idea that the strong excess emission observed in some Galactic
low-mass, pre-Main Sequence stars (T~Tauri stars) is produced by
accretion of material from a circumstellar disk dates back to the
pioneering work of Lynden-Bell \& Pringle (1974). The excess
luminosity is, then, related to the mass accretion rate. In
particular, the Balmer continuum radiation produced by the material
from the disk as it hits the stellar surface has been used as an
estimator of the mass infall activity (see, for example, Gullbring et
al 1998 and references therein).

First, we have identified candidate pre-Main Sequence stars in the
field of SN~1987A in the LMC through their Balmer continuum and
$\mathrm{H}\alpha$ excesses.  We have, then, derived the accretion
rate onto the central star with the following equations:

\begin{equation}
\left\{
\begin{array}{ll}
 L_{acc}\simeq\frac{GM_{\ast}\dot{M}}{R_{\ast}}\left(1-\frac{R_{\ast}}{R_{in}}
   \right)\\
 \log\left(\frac{L_{acc}}{L_\odot}\right)=1.16\ \log
   \left(\frac{L_{\mathrm{F336W},exc}}{L_\odot}\right)+1.24
\end{array}
\right.
\end{equation}
The second equation is the Gullbring et al (1998) empirical relation
between the accretion luminosity $L_{acc}$ and the Balmer excess
luminosity, as transformed by Robberto et al (2004) to the WFPC2 F336W
filter. The reader is referred to Romaniello et al (2004) for a thorough
description of the derivation of $\dot{M}$.

When interpreted as pre-Main Sequence stars, the comparison of the
objects' location in the HR diagram with theoretical evolutionary
tracks allows one to derive their masses ($\sim1-1.4~M_\odot$) and
ages ($\sim12-16$~Myrs). At such an age and with an accretion rate in
excess of $\sim 1.5\times10^{-8}M_\odot\ yr^{-1}$, these candidate
pre-Main Sequence stars in the field of SN~1987A are both older and
more active than their Galactic counterparts known to date. In fact,
the overwhelming majority of T~Tauri stars in Galactic associations
seem to dissipate their accretion disks before reaching an age of
about 6~Myrs (Haisch et al 2001; Armitage et al 2003). Moreover, the
oldest Classical T~Tauri star know in the Galaxy, TW~Hydr\ae, at an
age of 10~Myrs, \emph{i.e.} comparable to that of our sample stars,
has a measured accretion rate some 30 times lower than the stars in
the neighborhood of SN~1987A Muzerolle et al (2000).

The situation is summarized in Figure~1, where we compare the position
in the age-$\dot{M}$ plane of the stars described here with that of
members of Galactic star-forming regions. An obvious selection bias
that affects our census is that we only detect those stars with the
largest Balmer continuum excesses, \emph{i.e.} highest accretion
rates. There might be stars in the field with smaller accretion rates,
either intrinsically or because they were observed when the accretion
activity was at a minimum, which fall below our detection
threshold. This selection effect is rather hard to quantify, but it is
clear that the locus of the accreting stars that we do detect in the
neighborhood of SN~1987A is significantly displaced from the one
defined by local pre-Main Sequence stars.

\begin{figure}[!ht]
{\centerline{\psfig{file=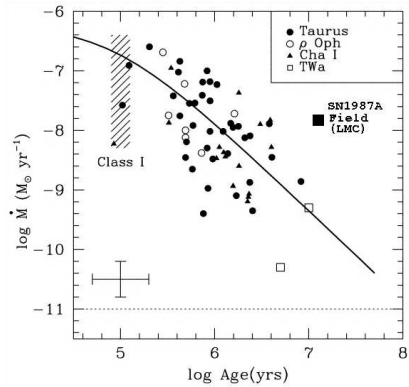,width=0.45\linewidth}}
\caption{Mass accretion rate as a function of age for Classical
   T~Tauri stars in different star-forming regions (adapted from
   Muzerolle et al 2000). Our result for the field of SN~1987A is
   marked with a black square.}}
\end{figure}


\begin{chapthebibliography}{}
\bibitem[Armitage et al (2003)]{arm03} Armitage, P.J., Clarke, C.J., and
  Palla, F. 2003, MNRAS, 342, 1139

\bibitem[Bessell (1991)]{bes91} Bessell, M.S. 1991, A\&A, 242, L17

\bibitem[Bonnell et al (2001)]{bon01} Bonnell, I.A., Clarke, C.J.,
 Bate, M.R., and Pringle, J. E. 2001, MNRAS, 324, 573

\bibitem[de Wit et al (2002)]{dewit02} de Wit, W. J., Beaulieu, J. P., and
  Lamers, H.J.G.L.M. 2002, A\&A, 395, 829

\bibitem[Gullbring et al (1998)]{gull98} Gullbring, E., Hartmann, L.,
  Brice\~{n}o, C., and Calvet, N. 1998, ApJ, 492, 323

\bibitem[Lamers et al (1999)]{lam99} Lamers, H.J.G.L.M., Beaulieu, J. P., and
  de Wit, W. J. 1999, A\&A, 341, 827

\bibitem[Lynden-Bell and Pringle (1974)]{lynd74} Lynden-Bell, D., and Pringle,
  J.E. 1974, MNRAS, 168, 603

\bibitem[Haisch et al (2001)]{hai01} Haisch, K.E., Jr., Lada, E.A, and Lada,
  C.J. 2001, ApJ, 553, L153

\bibitem[Hartman et al (1998)]{har98} Hartmann, L., Calvet, N., Gullbring, E.,
  and D'Alessio, P. 1998, ApJ, 494, 385

\bibitem[Hodge (1988)]{hod88}Hodge, P.W. 1988, PASP, 100, 1051

\bibitem[Muzerolle et al (2000)]{muz00} Muzerolle, J., Calvet, N.,
  Brice\~{n}o, C., Hartmann, L., and Hillenbrand, L. 2000, ApJ, 535, L47

\bibitem[Padget (1996)]{padg96} Padgett, D.L. 1996, AJ, 471, 874.

\bibitem[Palla and Stahler (1991)]{ps91} Palla, F., and Stahler, S. 1991,
  ApJ, 360, 47

\bibitem[Panagia et al (2000)]{pan00} Panagia, N., Romaniello, M., Scuderi, S.,
  and Kirschner, R.P. 2000, ApJ, 539, 197

\bibitem[Robberto et al (2004)]{rob04} Robberto, M., Song, J., Mora Carillo, G.
  Beckwith, S.V.W., Makidon, R.B., and Panagia, N. 2004, ApJ, 606, 952

\bibitem[Romaniello at al (2000)]{rom00} Romaniello, M., Salaris, M., Cassisi,
  S., and Panagia, N. 2000, APJ, 530, 738

\bibitem[Romaniello et al (2004)]{rom04} Romaniello, M., Robberto, M.,
  and Panagia, N. 2004, ApJ, 608, 220

\bibitem[van der Marel and Cioni (2001)]{mar01} van der Marel, R.P., and Cioni,
  M.-R.L. 2001, AJ, 122, 1807

\end{chapthebibliography}
\end{document}